\documentclass[1p]{elsarticle}
\usepackage{graphicx}
\usepackage{color}
\usepackage{amssymb}

\begin{document}
\title{Evidence for the $\psi (5S)$ and $\psi (4D)$
$c\bar{c}$ vector resonances}
\author[eef]{Eef van Beveren}
\ead{eef@teor.fis.uc.pt}
\author[george]{George Rupp}
\ead{george@ist.utl.pt}
\cortext[george]{Corresponding author}
\address[eef]{Centro de F\'{\i}sica Computacional,
Departamento de F\'{\i}sica, Universidade de Coimbra,
P-3004-516 Coimbra, Portugal}
\address[george]{Centro de F\'{\i}sica das Interac\c{c}\~{o}es Fundamentais,
Instituto Superior T\'{e}cnico, Edif\'{\i}cio Ci\^{e}ncia, Piso 3,
P-1049-001 Lisboa, Portugal}
\date{\today}

\begin{abstract}
We present evidence for the $\psi (5S)$ and $\psi (4D)$
$c\bar{c}$ vector resonances in experimental data
published by the Belle and BaBar Collaborations.
Central masses and resonance widths are estimated.
\end{abstract}


\maketitle

Since the late 1970s or early 1980s, it has been recognized
that the interpretation of strong-interaction scattering and
production data is much more involved than what is actually
being practised in data analysis. At present, one of the main
obstacles to steady progress in low-energy strong-interaction
physics is the very poor handling of threshold enhancements
in reactions where multihadron systems are produced.
Very often, the corresponding signals are not even considered
in data analysis, since their amplitudes are well below
some arbitrarily defined background.  Moreover, just above the
thresholds of the specific channels selected for analysis,
and where the threshold enhancements are usually well visible,
the amplitudes are fitted with simple Breit-Wigner shapes
and the associated central masses and widths. Thus, such
enhancements are --- by definition --- declared resonances.

In this respect, an important observation was made
by the BES Collaboration in Ref.~\cite{ARXIV08070494}.
To our knowledge, BES was the first to realize
that the $\psi (3770)$ cross section is built up
by two different amplitudes, viz.\ a relatively broad signal and a
rather narrow $c\bar{c}$ resonance.
For the narrow resonance, which probably corresponds to
the well-established $\psi (1D)(3770)$,
BES determined a central resonance position
of $3781.0\pm 1.3\pm 0.5$ MeV
and a width of $19.3\pm 3.1\pm 0.1$ MeV (their solution 2).
If the latter parameters are indeed confirmed,
it would be yet another observation
of a quark-antiquark resonance width
that is very different from the world average
($83.9\pm 2.4$ MeV \cite{PLB667p1} in this case),
after a similar result was obtained by the BaBar Collaboration
in Ref.~\cite{PRL102p012001}, for $b\bar{b}$ resonances.
Concerning the broader structure, the BES Collaboration indicated,
for their solution 2,
a central resonance position of $3762.6\pm 11.8\pm 0.5$ MeV
and a width of $49.9\pm 32.1\pm 0.1$ MeV.
The signal significance for this new enhancement is $7.6\sigma$
(solution 2).
It was explained as a possible diresonance
\cite{PRD78p116014} or heavy molecular state \cite{ARXIV08080073}.

Moreover, in the latter BES publication,
the existence of conflicting results
with respect to the branching fraction
for non-$D\bar{D}$ hadronic decays
of the $\psi (1D)(3770)$ was emphasized.
On the one hand, the total branching fraction
for exclusive non-$D\bar{D}$ modes has been measured
to be less than 2\% \cite{PLB605p63,PRD74p012005}.
But on the other hand, for inclusive non-$D\bar{D}$ decay modes,
values of about 15\% have been found \cite{PRD76p122002,PLB659p74}.
According to BES,
this apparent discrepancy may partially be due
to the assumption that the line shape above the $D\bar{D}$ threshold
is the result of one simple resonance.
Now, in Ref.~\cite{PRD80p074001} we have shown
that the broader structure is most likely
caused by a non-resonant contribution to the production amplitude,
thus lending further support to the idea that
the $\psi (1D)(3770)$ enhancement consists of two distinct signals,
one of which is nothing else but a threshold effect.

Several of the new enhancements in production cross sections
share the common property
that they occur at --- or just above --- an important threshold.
The most recent example, viz.\ the $J/\psi\,\phi$ enhancement
observed and baptized as $Y(4140)$ by the CDF Collaboration
\cite{PRL102p242002},
appears right above the $J/\psi\,\phi$ threshold.
The enhancement in
the $e^{+}e^{-}\to\Lambda_{c}^{+}\Lambda_{c}^{-}$ cross section,
reported by Belle \cite{PRL101p172001},
occurs right above the $\Lambda_{c}^{+}\Lambda_{c}^{-}$ threshold.
The $Y(4260)$ enhancement in
the $e^{+}e^{-}\to J/\psi\pi^{+}\pi^{-}$ cross section,
observed by BaBar \cite{PRL95p142001},
is right on top of the $D_{s}^{\ast}D_{s}^{\ast}$ threshold.
The $X(3872)$ enhancement \cite{PRL91p262001} in
$B^{\pm}\to J/\psi K^{\pm}\pi^{+}\pi^{-}$ decay
lies just above the $DD^{\ast}$ threshold.

\begin{figure}[htbp]
\begin{center}
\begin{tabular}{c}
\includegraphics[width=200pt]{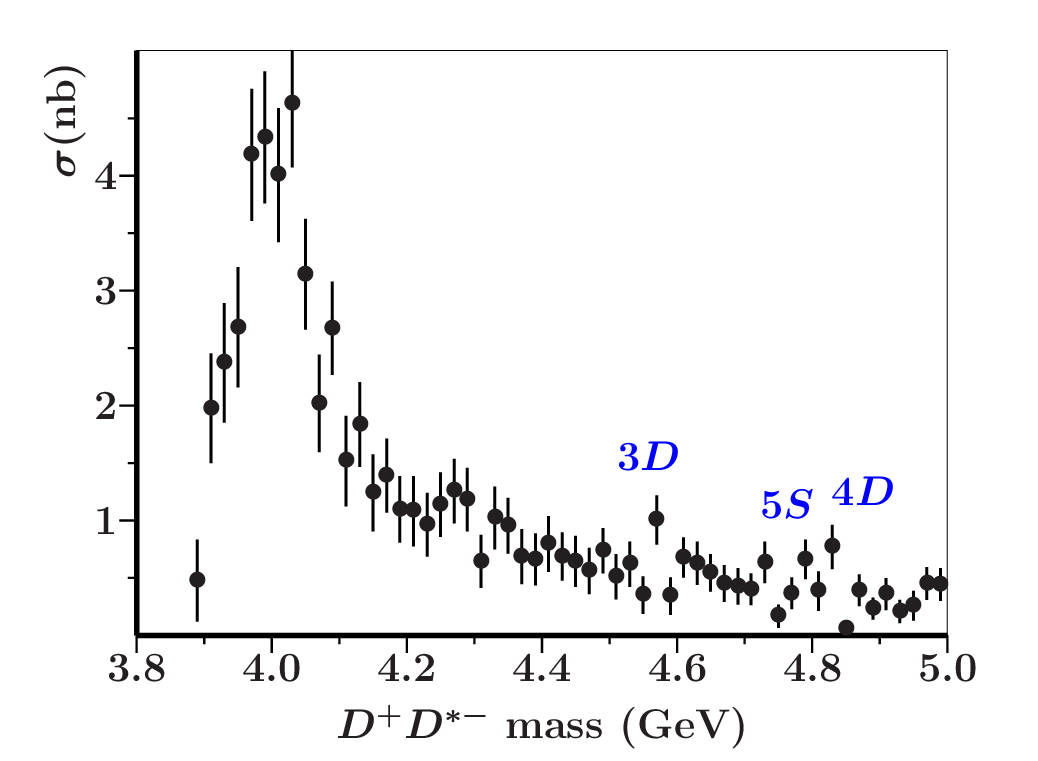}\\ [-20pt]
\end{tabular}
\end{center}
\caption{\small
Event distributions for the reaction
$e^{+}e^{-}\to D^{+}\bar{D}^{\ast -}$,
as published by the Belle Collaboration
\cite{PRL98p092001}.
}
\label{3D5Sand4D}
\end{figure}

In Fig.~\ref{3D5Sand4D}, we show the production cross sections
for the reaction $e^{+}e^{-}\to D^{+}\bar{D}^{\ast -}$,
published by the Belle Collaboration \cite{PRL98p092001},
using initial-state radiation (ISR).
The signal was not further analysed by Belle,
for which they gave the following reason:
\begin{quote}
{\it Since a reliable fit to the cross sections
[obtained above]
requires a solution to a non-trivial and
model-dependent problem of coupled channels and threshold effects, we
do not report results here.}
\end{quote}
This is symptomatic for the quite desperate situation
in which low-energy strong-interac\-tion physics
finds itself at present.
We have no doubts that the major part of the amplitude
at about 4.0 GeV is due to threshold enhancements.
However, contrary to the single threshold enhancement
under the $\psi (3770)$,
here the amplitude contains several enhancements,
viz.\
$D^{\pm}D^{\ast\mp}$ at 3.880 GeV,
$D_{s}^{\pm}D_{s}^{\mp}$ at 3.937 GeV,
$D^{\ast\pm}D^{\ast\mp}$ at 4.02 GeV
and $D_{s}^{\pm}D_{s}^{\ast\mp}$ at 4.081 GeV.
It will certainly not be a simple task to conveniently parametrize
these mutually interfering threshold enhancements,
which in their turn interfere with the resonances
of the quark-antiquark propagator.

In Ref.~\cite{IJTPGTNO11p179}, we derived a precise relation between the
formalism of non-exotic meson-meson scattering due to a resonating
$s$-channel quark-antiquark propagator in the intermediate state, and the
deformed $q\bar{q}$ resonance spectrum owing to the inclusion of infinite
chains of meson loops.
Moreover, in Ref.~\cite{AP323p1215} we deduced an amplitude for production
processes, resulting in a complex relation \cite{EPL81p61002}
between production and scattering amplitudes.
The latter relation is formally equivalent \cite{EPL84p51001}
to the real relation of Au, Morgan, and Pennington \cite{PRD35p1633},
but with an important difference:
whereas the coefficients of the complex relation \cite{EPL81p61002}
are of a purely kinematical origin,
the real coefficients of Ref.~\cite{PRD35p1633}
contain the scattering amplitudes themselves
\cite{EPL84p51002}.
As a consequence, one does not find
a distinct threshold enhancement
in the formalism of Ref.~\cite{PRD35p1633}.

Furthermore, in Ref.~\cite{ARXIV09100967}
we studied the shapes of open-bottom thresholds.
However, we must admit that the case
of a clear separation of
$b\bar{b}\leftrightarrow (b\bar{u}/\bar{d})+(\bar{b}u/d)$
and
$b\bar{b}\leftrightarrow (b\bar{s})+(\bar{b}s)$
thresholds
is much simpler than the comparable case of open charm,
in which the corresponding channels are partly overlapping, as
$m_D<m_{D_s}<m_{D^\ast}$, whereas $m_B<m_{B^\ast}<m_{B_s}$.
Consequently, practical expressions for data analysis are not yet at hand.

We can observe some structure in the data \cite{PRL98p092001} presented in
Fig.~\ref{3D5Sand4D}, at invariant masses in the interval 4.7--4.9 GeV,
where the $\psi (5S)$ and $\psi (4D)$ $c\bar{c}$ vector resonances
are expected \cite{PRD21p772} to reside,
besides an enhancement in the $\psi (3D)$ region.
Here, despite our still very incomplete description
of threshold enhancements, we shall continue our program to search for
new vector $c\bar{c}$ states in the data, and indicate where to look.
For now, we shall concentrate on
the $\psi (5S)$ and $\psi (4D)$ resonances.

In Ref.~\cite{PRL101p172001}, the Belle Collaboration announced
the observation of a near-threshold enhancement,
by studying the $e^{+}e^{-}\to\Lambda_{c}^{+}\Lambda_{c}^{-}$ cross section.
The experimental analysis resulted in a mass and width
for this enhancement of $M=(4634^{+8}_{-7})$(stat.)$^{+5}_{-8}$(sys.) MeV
and $\Gamma_{\mathrm{tot}}=92^{+40}_{-24}$(stat.)$^{+10}_{-21}$(sys.) MeV,
respectively \cite{PRL101p172001}, with a significance of $8.8$ $\sigma$.
An intriguing aspect of this experimental observation
is that the main signal lies close to
the $\Lambda_{c}^{+}\Lambda_{c}^{-}$ threshold,
making an understanding of this structure a highly topical issue
\cite{EPL85p61002}.

\begin{figure}[htbp]
\begin{minipage}[b]{0.48\linewidth}
\centering
\includegraphics[width=170pt]{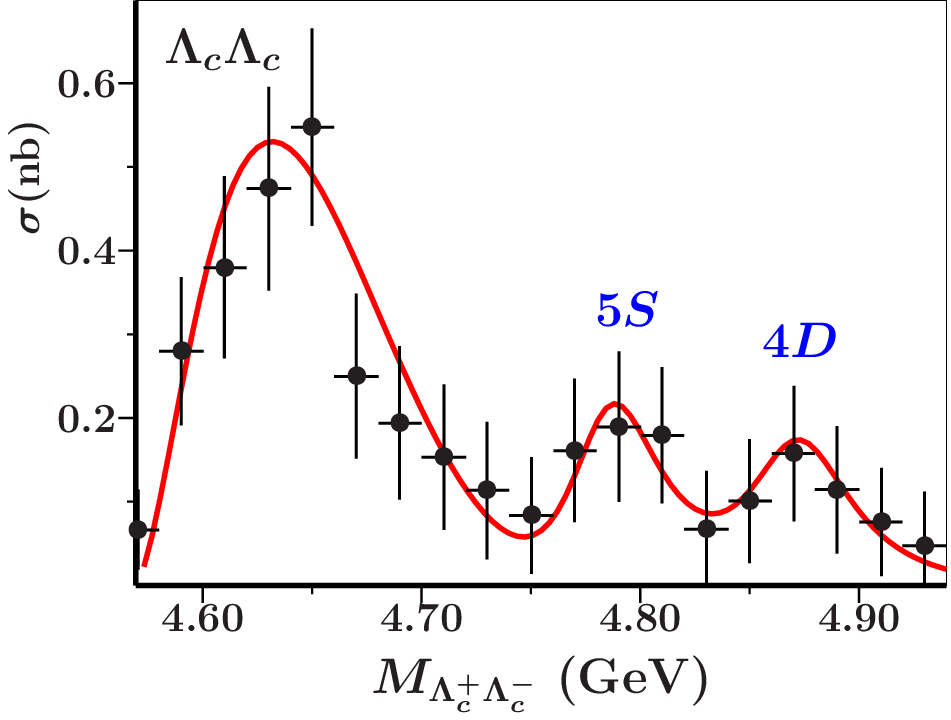}
\caption{\small
Event distributions for the reaction
$e^{+}e^{-}\to\Lambda_{c}^{+}\Lambda_{c}^{-}$,
as published by the Belle Collaboration
in Ref.~\cite{PRL101p172001}.}
\label{LcLc5Sand4D}
\end{minipage}
\hspace{0.02\linewidth}
\begin{minipage}[b]{0.48\linewidth}
\centering
\includegraphics[width=170pt]{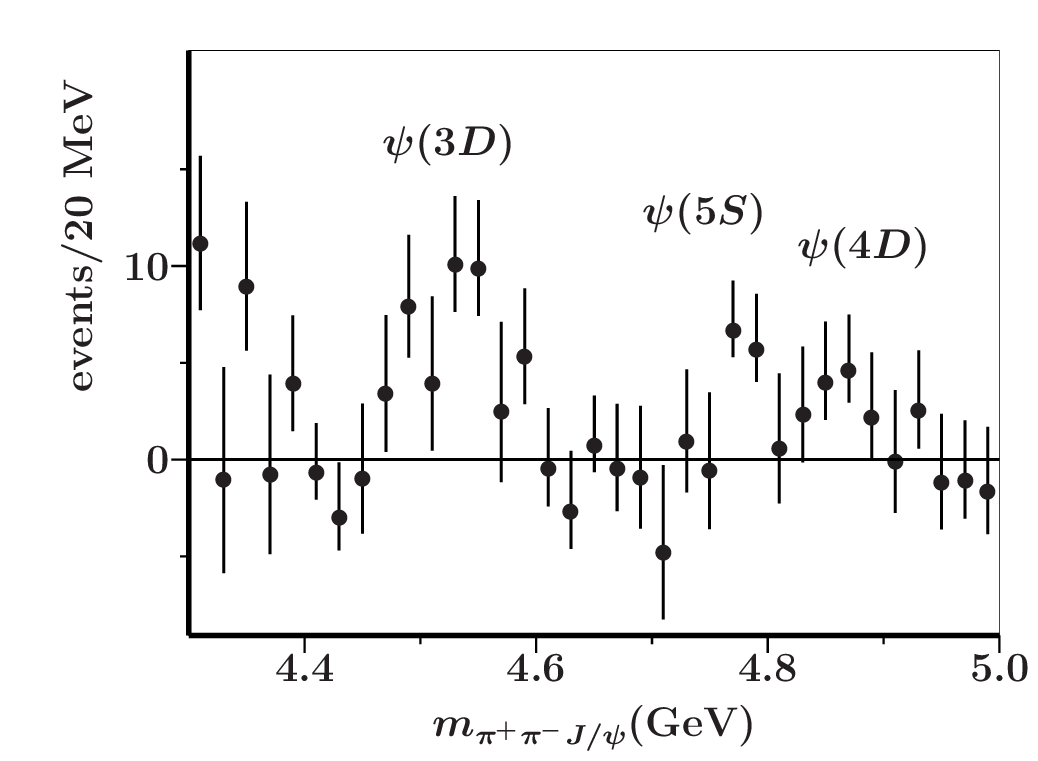}
\caption{\small
Event distributions for the missing signal \cite{ARXIV09044351}
in the reaction $e^{+}e^{-}\to\pi^{+}\pi^{-}J/\psi$,
using data published by the BaBar Collaboration
\cite{PRL95p142001}.}
\label{E4260}
\end{minipage}
\end{figure}

The $e^{+}e^{-}\to\Lambda_{c}^{+}\Lambda_{c}^{-}$ cross section
is given in Fig.~\ref{LcLc5Sand4D},
where one observes, besides the threshold enhancement,
two well separated resonances, i.e., most probably the
$\psi (5S)$ and $\psi (4D)$ $c\bar{c}$ vector resonances
as discussed in Ref.~\cite{EPL85p61002}.

Modelling the $\Lambda_{c}^{+}\Lambda_{c}^{-}$
enhancement might of course be done by considering
a full wave function with all possible components, viz.\
$c\bar{c}$ states, charmed-meson pairs, and charmed-baryon pairs.
Such a wave function will have a large
$\Lambda_{c}^{+}\Lambda_{c}^{-}$ component under the enhancement.
Nevertheless, it will not give rise to a resonance pole
in the full coupled-channel scattering amplitude,
unless, accidentally, there is a dynamically generated
resonance pole in this invariant-mass region.
Now, in view of the large non-resonant contribution
to the $\Lambda_{c}^{+}\Lambda_{c}^{-}$ enhancement,
we do not consider the occurrence here of a dynamically generated
resonance very likely,
which is in line with a similar conclusion by Bugg
\cite{JPG36p075002}.
Recently, it was claimed \cite{ARXIV10052055} that the
enhancement could be consistent
with the $\psi (2S)f_0$(980) molecular picture of the $Y$(4660),
taking into account the $\Lambda_{c}^{+}\Lambda_{c}^{-}$ final-state
interaction.
In our opinion, it certainly does not represent
a state of the $J^{PC}=1^{--}$ $c\bar{c}$ spectrum
as advocated in Refs.~\cite{PRD78p114033,PRD79p094004}.

Earlier, however,
the BaBar Collabaration had published \cite{PRL95p142001} interesting data
for the reaction $e^{+}e^{-}\to\pi^{+}\pi^{-}J/\psi$, using ISR.
These data, which passed unnoticed for a few years, showed
that, upon studying the missing signal \cite{ARXIV10044368},
one could observe (see Fig.~\ref{E4260})
the $\psi (5S)$ and $\psi (4D)$ $c\bar{c}$ vector resonances,
besides a clear signal from the $\psi (3D)$ \cite{ARXIV09044351}.

More recent data from Belle and BaBar reveal clearer
resonance shapes for the $\psi (5S)$ and $\psi (4D)$.
Here, we study data for the invariant-mass distributions
of the $D^{0}D^{\ast -}\pi^{+}$ and $D^{0}D^{-}\pi^{+}$ systems,
published by Belle in Refs.~\cite{PRD80p091101}
and \cite{PRL100p062001}, respectively,
and also of the $D^{\ast}\bar{D}^{\ast}$ system,
published by BaBar \cite{PRD79p092001}.

\begin{figure}[htbp]
\begin{minipage}[b]{0.48\linewidth}
\centering
\includegraphics[width=170pt]{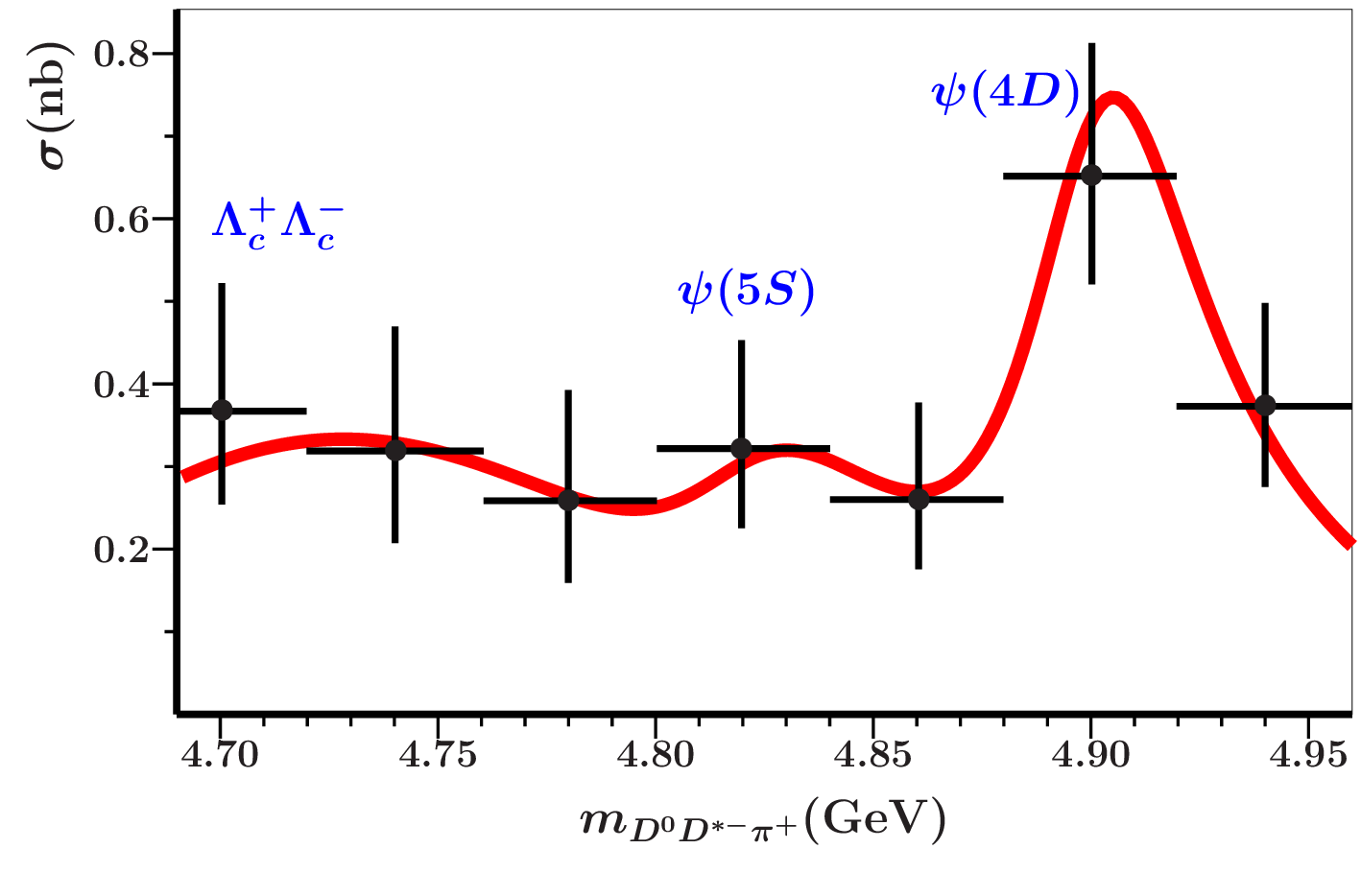}
\caption{\small
Fit to the $D^{0}D^{\ast -}\pi^{+}$ invariant-mass distribution for
the Belle data of Ref.~\cite{PRD80p091101}.}
\label{D0Dsmpip}
\end{minipage}
\hspace{0.02\linewidth}
\begin{minipage}[b]{0.48\linewidth}
\centering
\includegraphics[width=170pt]{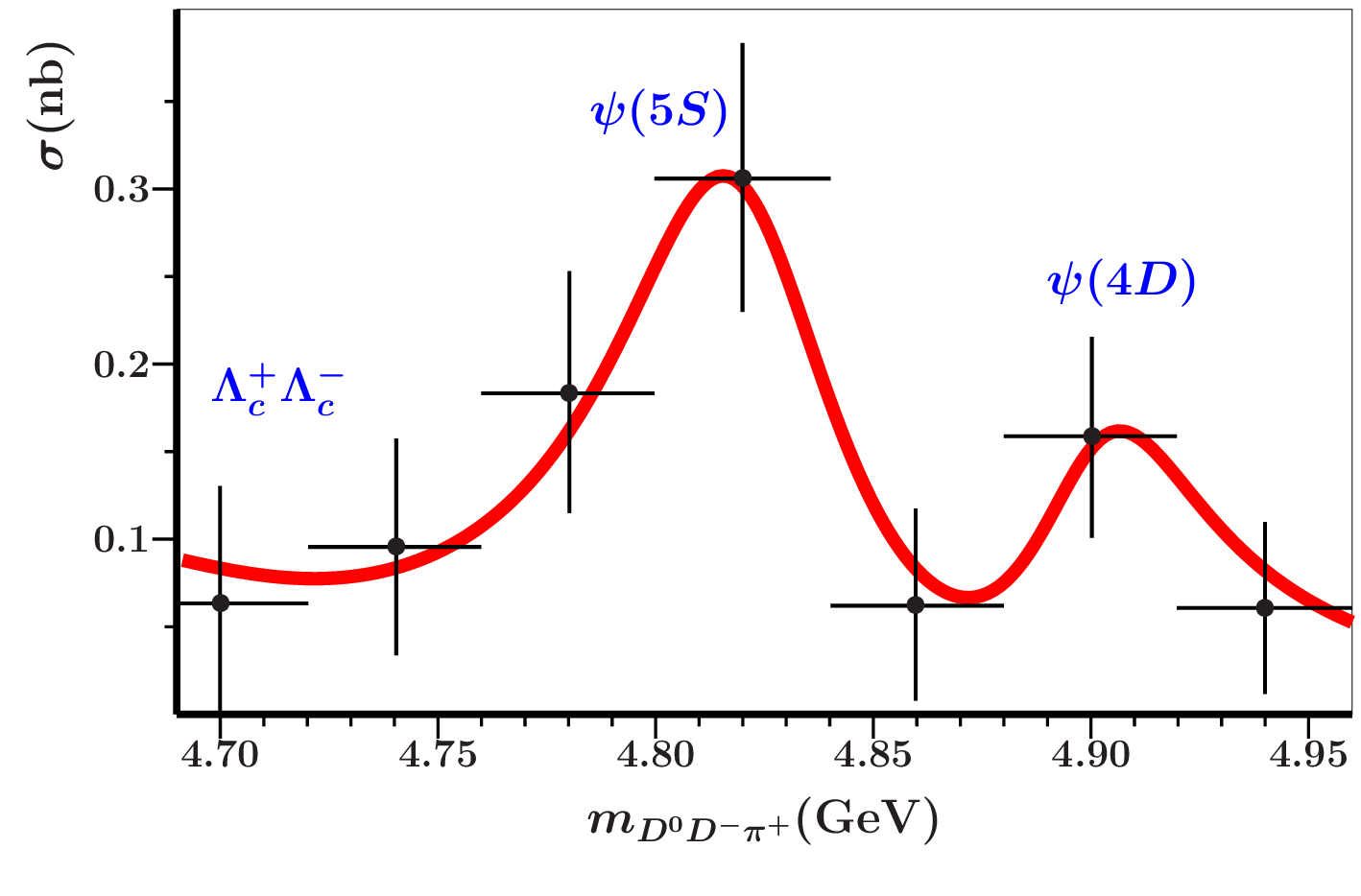}
\caption{\small
Fit to the  $D^{0}D^{-}\pi^{+}$ invariant-mass distribution for
the Belle data of Ref.~\cite{PRL100p062001}.}
\label{D0Dmpip}
\end{minipage}
\end{figure}

\begin{figure}[htbp]
\begin{center}
\begin{tabular}{c}
\includegraphics[width=180pt]{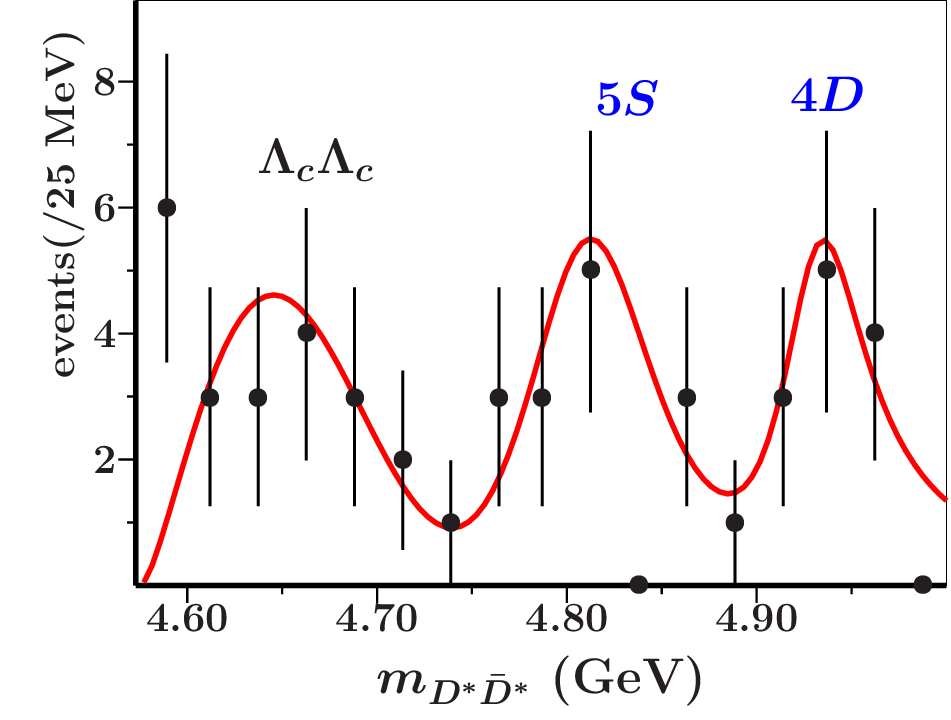}\\ [-20pt]
\end{tabular}
\end{center}
\caption{\small
Fit to the $D^{\ast}\bar{D}^{\ast}$ invariant-mass distribution for
the BaBar data of Ref.~\cite{PRD79p092001}.
}
\label{DstarDstar}
\end{figure}

In Figs.~\ref{D0Dsmpip} and \ref{D0Dmpip},
we show our fit to the relevant data
of the $D^{0}D^{\ast -}\pi^{+}$ \cite{PRD80p091101}
and $D^{0}D^{-}\pi^{+}$ \cite{PRL100p062001}
invariant-mass distributions, respectively ,
for $M(5S)=4.82$ GeV, $M(4D)=4.90$ GeV,
$\Gamma (5S)=65$ MeV, and $\Gamma (4D)=50$ MeV, in both cases.
Errors in these values will be of the order of the bin sizes,
viz.\ 40 MeV.
In Fig.~\ref{DstarDstar},
we present our fit to the relevant data of the
$D^{\ast}\bar{D}^{\ast}$invariant-mass distribution \cite{PRD79p092001},
for $M(5S)=4.81$ GeV, $M(4D)=4.93$ GeV,
$\Gamma (5S)=100$ MeV, and $\Gamma (4D)=55$ MeV
(bin sizes are 25 MeV here).

Previously \cite{PRD80p074001}, we had found
$M(5S)=4.784$ GeV, $M(4D)=4.871$ GeV,
$\Gamma (5S)=55$ MeV, and $\Gamma (4D)=60$ MeV
(bin sizes of 20 MeV)
for the Breit-Wigner parameters
of the $\psi (5S)$ and $\psi (4D)$,
in order to fit the $\Lambda_{c}^{+}\Lambda_{c}^{-}$
mass distribution of Fig.~\ref{E4260}.

The latter results merely demonstrate that Breit-Wigner parameters
are only useful for narrow resonances,
which in all channels have approximately the same appearance.
For many strong processes, resonance poles
can only be determined through analytic continuation of the scattering
amplitude obtained from multichannel coupled equations.
As a consequence, it does not make much sense either
to average over the various Breit-Wigner parameters we have obtained so far
for the $\psi (5S)$ and $\psi (4D)$ resonances.

In conclusion, we can state that we have found clear
signs of the $\psi (5S)$ and $\psi (4D)$ $c\bar{c}$ vector resonances in
published data. Depending upon the channel in which these new states are
analysed, they will show up with central masses of
$M(5S)\approx$~4.78--4.81 GeV and $M(4D)\approx$~4.87-4.93 GeV,
while their widths will be in the ball park of
$\Gamma (5S)\approx$~50--100 MeV and $\Gamma(4D)\approx$~50--60 MeV.
\vspace{0.3cm}

{\bf Acknowledgements}:
We are grateful for the precise measurements
of the Belle and BaBar Collaborations, which
made the present analysis possible.
This work was supported in part by the {\it Funda\c{c}\~{a}o para a
Ci\^{e}ncia e a Tecnologia} \/of the {\it Minist\'{e}rio da Ci\^{e}ncia,
Tecnologia e Ensino Superior} \/of Portugal, under contract
CERN/\-FP/\-109307/ 2009.

\newcommand{\pubprt}[4]{#1 {\bf #2}, #3 (#4)}
\newcommand{\ertbid}[4]{[Erratum-ibid.~#1 {\bf #2}, #3 (#4)]}
\def\AP{Ann.\ Phys.}
\def\EPL{Europhys.\ Lett.}
\def\IJTPGTNO{Int.\ J.\ Theor.\ Phys.\ Group Theor.\ Nonlin.\ Opt.}
\def\JPG{J.\ Phys.\ G}
\def\PLB{Phys.\ Lett.\ B}
\def\PRD{Phys.\ Rev.\ D}
\def\PRL{Phys.\ Rev.\ Lett.}

\end{document}